*mathematics*

MDPI*Article*

# Collaboration Effect by Co-Authorship on Academic Citation and Social Attention of Research

Pablo Dorta-González [1,*] and María Isabel Dorta-González [2]1 Institute of Tourism and Sustainable Economic Development (TIDES), Campus de Tafira, University of Las Palmas de Gran Canaria, 35017 Las Palmas de Gran Canaria, Spain
2 Department of Ingeniería Informática y de Sistemas, University of La Laguna, Avenida Astrofísico Francisco Sánchez s/n, 38271 La Laguna, Spain; isadorta@ull.es
* Correspondence: pablo.dorta@ulpgc.es**Abstract:** Academic citation and social attention measure different dimensions of the impact of research results. Both measures do not correlate with each other, and they are influenced by many factors. Among these factors are the field of research, the type of access, and co-authorship. In this study, the increase in the impact due to co-authorship in scientific articles disaggregated by field of research and access type, was quantified. For this, the citations and social attention accumulated until the year 2021 by a total of 244,880 research articles published in the year 2018, were analyzed. The data source was *Dimensions.ai*, and the units of study were research articles in Economics, History and Archaeology, and Mathematics. As the main results, a small proportion of the articles received a large part of the citations and most of the social attention. Both citations and social attention increased, in general, with the number of co-authors. Thus, the greater the number of co-authors, the greater the probability of being cited in academic articles and mentioned on social media. The advantage in citation and social attention due to collaboration is independent of the access type for the publication. Furthermore, although collaboration with an additional co-author is in general positive in terms of citation and social attention, these positive effects reduce as the number of co-authors increases.

**Keywords:** collaboration; co-authorship; citation; social attention; altmetric; open access

**MSC:** 62P25**Citation:** Dorta-González, P.; Dorta-González, M.I. Collaboration Effect by Co-Authorship on Academic Citation and Social Attention of Research. *Mathematics* **2022**, *10*, x. https://doi.org/10.3390/xxxxx

Academic Editors: Francisco Chiclana, Sergei Petrovskii, Matjaž Perc, Antonio Di Crescenzo and Marjan Mernik

Received: date
Accepted: date
Published: date

**Publisher's Note:** MDPI stays neutral with regard to jurisdictional claims in published maps and institutional affiliations.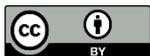

**Copyright:** © 2022 by the authors. Submitted for possible open access publication under the terms and conditions of the Creative Commons Attribution (CC BY) license (https://creativecommons.org/licenses/by/4.0/).## 1. Introduction

Collaboration is increasingly important to progress in scientific research. Research collaboration is fundamental because it links knowledge and competencies into novel ideas and research lines [1].

There is a debate about what constitutes research collaboration [2]. However, a verifiable way to approach collaboration is to analyze the co-authorship of research papers [3]. So, co-authorship constitutes a proxy for research collaboration [4,5].

The production of research articles by researchers has increased, particularly in recent decades. However, the number of co-authors has also increased. Thus, adjusted for collaboration, the publication rate of researchers in all research fields has not increased overall [6].

The counting of research articles is strongly and significantly associated with the number of co-authors [7,8]. However, when examining the production by fractional counting, i.e., dividing the number of papers by the number of collaborators, the number of co-authors is not a significant predictor of productivity [8].

Collaboration has been related with more citations and higher-impact science [7]. Furthermore, more collaborative scientists tend to have higher h-indices [9]. However,

*Mathematics* **2022**, *10*, x. https://doi.org/10.3390/xxxxx www.mdpi.com/journal/mathematics

this effect varies with research fields. For instance, in biology, earth sciences, and social sciences, the relationship is not significant. However, there is a strong relationship in physics [9]. These authors indicate that the strong association in physics is a result of the growing number of large, high impact, intensely collaborative projects in experimental physics.

Considering publications, increased collaboration has been related with the highest citations [10]. In the field of information systems, the total number of collaborators is positively associated with citations received [11]. This positive effect of collaboration on citations exists also across disciplines in Italy [12], as well as for economists [13] and biologists [14]. However, there exists controversy about the real effect of open access modalities on the visibility and citation impact of publications [15–17].

Governments increasingly push researchers toward activities with social impact, including economic, cultural and health benefits [18]. Since the term 'altmetrics' was introduced in 2010 [19], theoretical and practical research has been carried out in this discipline [20].

One advantage of alternative metrics compared to citations is that altmetrics tend to accumulate scores in the first few months after publication [21]. However, except for the Mendeley readership count, which is moderately correlated with citations [22], there is a negligible or weak correlation between citations and most altmetric indicators at the publication level [23,24]. This means that altmetrics might capture diverse forms of impact, which are different from the citation impact [25].

In this paper, we quantify the contribution of collaboration to academic citations and social attention of research. As a novel methodology, we measure the contribution due to co-authorship through a marginal analysis. This is, through the contribution to the impact, in citation and social attention, attributable to an additional co-author. We also analyzed for the first time the possible effect of the access type of the publication, particularly for open access modalities.

For this purpose, we considered the accumulated citations and social attention until the year 2021 of a total of 244,880 research articles published in the year 2018. The data source is *Dimensions.ai*, and the research articles analyzed correspond to those classified in this database in the research fields of Economics, History and Archaeology, and Mathematics.

## 2. Materials and Methods

This paper used the number of co-authors as a proxy of the intensity of collaboration. This assumption has been used since the 1960s [3,5,26,27].

The unit of study is the research article, and the source of data is *Dimensions.ai*. This database classifies research articles into 22 fields of research. The articles analyzed in this paper correspond to those classified in this database in the following three research fields: Economics, History and Archaeology, and Mathematics. Note that each of these fields belongs to a different branch of knowledge.

The source of altmetric data in *Dimensions.ai* is *Altmetric.com (accessed on 15 March 2022)*. This is a currently popular and one of the first altmetric aggregator platforms, that originated in 2011 with the support of *Digital Science*. It tracks and accumulates mentions and views from different social media channels, news, blogs, and other platforms for scholarly articles. It also computes a weighted score called the 'altmetric attention score' where each citation category contributes differently to the final score [28].

The altmetric attention score measures the social attention a paper receives from mainstream and social media, public policy documents, *Wikipedia*, etc. It collects the online presence, and analyzes the conversations around the research. For clarity in results, this measure is referred to as the social attention score in this paper, or just social attention.

*Dimensions.ai* is currently one of the most extensive databases of bibliographic references and with the greatest coverage of regional/local journals. Note that *Google Scholar* is not strictly a database, but rather a search engine for academic documents that links to the websites it indexes. As indicated, *Dimensions.ai* covers a significant volume of regional and local journals, unlike other databases such as *Web of Science* and *Scopus*, in addition to the top journals that tend to have an international focus. Globally, there are numerous local journals, in many cases publishing in languages other than English. These journals are often open access and are not collected in databases such as *Web of Science* or *Scopus*. On the other hand, there are articles published in open access mega journals such as *PLoS ONE*, *Scientific Reports*, *Nature Communications*, *Sage Open*, etc.

In this paper, we considered all research articles indexed in the database in 2018, and citations and altmetric attention scores counted in the period of 2018–2021. Thus, a total of 244,880 research articles were analyzed. Data were exported on 15 March 2022.

## 3. Results

Citation and social mention do not correlate with each other (Table 1). The Pearson linear correlation coefficients between times cited and social attention were all positive, although they were quite low for the most prevalent co-authorships. Moreover, they vary considerably by the field of research and number of co-authors.

**Table 1.** Pearson correlation coefficient between times cited and social attention score in relation to co-authorship.

| Number of Authors | Economics | | History & Archaeology | | Mathematics | |
|---|---|---|---|---|---|---|
| | N | Coeff. | N | Coeff. | N | Coeff. |
| 1 | 17,072 | 0.38 | 48,620 | 0.32 | 25,147 | 0.13 |
| 2 | 16,263 | 0.21 | 5988 | 0.34 | 37,608 | 0.06 |
| 3 | 12,466 | 0.29 | 2399 | 0.29 | 30,888 | 0.21 |
| 4 | 5676 | 0.26 | 1336 | 0.22 | 17,307 | 0.13 |
| 5 | 2615 | 0.18 | 772 | 0.39 | 7815 | 0.15 |
| 6 | 1376 | 0.24 | 448 * | 0.69 | 3766 | 0.01 |
| 7 | 766 | 0.26 | 276 * | 0.68 | 1739 | 0.22 |
| 8 | 496 * | 0.75 | 186 * | 0.47 | 1010 * | 0.46 |
| 9 | 199 * | 0.33 | 109 * | 0.76 | 595 * | 0.11 |
| ≥10 | 528 * | 0.65 | 99 * | 0.59 | 1315 | 0.39 |
| Total | 57,457 | 0.27 | 60,233 | 0.33 | 127,190 | 0.12 |

* Prevalence < 1% (N < 1% of Total).

The lowest correlations were reached in Mathematics, of less than 0.22 for co-authorships with a prevalence of greater than 1%. The only correlations higher than 0.65 were reached in History and Archaeology and Economics, but for an unusually high number of authors in the standards of these fields. That is, from six authors in History and Archaeology, and from eight authors in Economics. However, the proportion of articles in these cases is very small and does not reach 1% of the total in the field. In cases of co-authorship with a prevalence of greater than 1% in History and Archaeology and Economics, the correlations are less than 0.39.

In terms of the prevalence in the number of co-authors (Table 2), in History and Archaeology, the most frequent is a single author, constituting 81% of the total articles in this field. In Economics, the most frequent is also a single author, although with a much lower prevalence (30%), while 80% of the articles have less than three authors. In Mathematics, the most frequent is two authors (30%), while 87% of the articles have less than four authors.

Given the differences found, we next analyzed the co-authorship effect on both impact measures according to each field (Table 2). For those co-authorships with a prevalence of greater than 1%, the average citations in Economics and Mathematics were of a similar order of magnitude, but much higher than the average citations in History and Archaeology.

Table 2. Times cited and social attention score in relation to co-authorship.

| Field of Research | Number of Authors *n* | Articles | | Times Cited | | | Social Attention Score | | |
|---|---|---|---|---|---|---|---|---|---|
| | | Number N | Prevalence | Mean | Mean Per Author | Marginal Contribution [1] | Mean | Mean per Author | Marginal Contribution [1] |
| | | | | Economics | | | | | |
| | 1 | 17,072 | 29.71% | 3.49 | 3.49 | - | 2.87 | 2.87 | - |
| | 2 | 16,263 | 28.30% | 6.83 | 3.42 | 96.0% | 4.36 | 2.18 | 51.5% |
| | 3 | 12,466 | 21.70% | 8.68 | 2.89 | 27.0% | 5.40 | 1.80 | 23.9% |
| | 4 | 5676 | 9.88% | 10.39 | 2.60 | 19.8% | 4.95 | 1.24 | −8.3% |
| | 5 | 2615 | 4.55% | 9.95 | 1.99 | −4.2% | 6.13 | 1.23 | 23.8% |
| | 6 | 1376 | 2.39% | 9.52 | 1.59 | −4.3% | 4.83 | 0.80 | −21.2% |
| | 7 | 766 | 1.33% | 11.28 | 1.61 | 18.4% | 6.92 | 0.99 | 43.4% |
| | 8 * | 496 | 0.86% | 11.64 | 1.46 | 3.3% | 12.21 | 1.53 | 76.3% |
| | 9 * | 199 | 0.35% | 7.24 | 0.80 | −37.9% | 4.25 | 0.47 | −65.2% |
| | ≥10 * | 528 | 0.92% | 16.58 | - | 129.1% | 17.92 | - | 321.5% |
| | Total | 57,457 | | | | | | | |
| | | | | History & Archaeology | | | | | |
| | 1 | 48,620 | 80.72% | 0.58 | 0.58 | - | 1.28 | 1.28 | - |
| | 2 | 5988 | 9.94% | 2.02 | 1.01 | 250.9% | 3.29 | 1.65 | 156.6% |
| | 3 | 2399 | 3.98% | 3.29 | 1.10 | 62.6% | 7.30 | 2.43 | 121.6% |
| | 4 | 1336 | 2.22% | 4.66 | 1.16 | 41.5% | 8.91 | 2.23 | 22.1% |
| | 5 | 772 | 1.28% | 5.91 | 1.18 | 27.0% | 13.91 | 2.78 | 56.1% |
| | 6 * | 448 | 0.74% | 7.15 | 1.19 | 21.0% | 28.68 | 4.78 | 106.2% |
| | 7 * | 276 | 0.46% | 9.12 | 1.30 | 27.5% | 16.40 | 2.34 | −42.8% |
| | 8 * | 186 | 0.31% | 9.43 | 1.18 | 3.4% | 26.20 | 3.28 | 59.8% |
| | 9 * | 109 | 0.18% | 10.96 | 1.22 | 16.3% | 28.71 | 3.19 | 9.5% |
| | ≥10 * | 99 | 0.16% | 12.04 | - | 9.8% | 56.24 | - | 95.9% |
| | Total | 60,233 | | | | | | | |
| | | | | Mathematics | | | | | |
| | 1 | 25,147 | 19.77% | 3.46 | 3.46 | - | 0.83 | 0.83 | - |
| | 2 | 37,608 | 29.57% | 6.00 | 3.00 | 73.3% | 0.88 | 0.44 | 6.0% |
| | 3 | 30,888 | 24.28% | 7.91 | 2.64 | 31.8% | 1.13 | 0.38 | 27.6% |
| | 4 | 17,307 | 13.61% | 9.87 | 2.47 | 24.9% | 1.19 | 0.30 | 5.4% |
| | 5 | 7815 | 6.14% | 11.42 | 2.28 | 15.6% | 1.77 | 0.35 | 48.8% |
| | 6 | 3766 | 2.96% | 10.69 | 1.78 | −6.4% | 1.82 | 0.30 | 2.7% |
| | 7 | 1739 | 1.37% | 11.57 | 1.65 | 8.3% | 3.04 | 0.43 | 67.3% |
| | 8 * | 1010 | 0.79% | 11.26 | 1.41 | −2.7% | 3.31 | 0.41 | 9.0% |
| | 9 * | 595 | 0.47% | 13.10 | 1.46 | 16.4% | 4.46 | 0.50 | 34.7% |
| | ≥10 | 1315 | 1.03% | 17.55 | - | 34.0% | 10.48 | - | 135.0% |
| | Total | 127,190 | | | | | | | |

* Prevalence < 1% (N < 1% of Total). [1] Marginal_contribution $(n) = (Mean\ (n) - Mean\ (n-1))/(Mean\ (n-1)) \times 100$, $n > 1$.

The social attention in Economics for one and two authors is greater than in History and Archaeology, although from three authors onwards this relationship is reversed.

Furthermore, research in Mathematics is by far the one that generates the least social attention.

It is observed in Table 2, for the most prevalent co-authorships in the three fields analyzed, that the mean in citation and social attention increases with the number of co-authors up to a certain maximum value. For co-authorships with a prevalence of greater than 1%, maximum values in citation are obtained for four authors in Economics, and five in History and Archaeology, and Mathematics. However, in terms of social attention, the maximum values in the mean score are obtained for three authors in Economics, five in History and Archaeology, and seven in Mathematics.

Therefore, comparative advantages are obtained in the impact in Economics at up to four authors in citation and three in social attention. In History and Archaeology, comparative advantages are obtained in both measures for up to five authors, after which collaborations cease to be frequent. In Mathematics, comparative advantages are obtained in the impact of up to five authors in citation and seven in social attention.

Considering the prevalence of co-authorship, citations increase with the number of co-authors in the vast majority of cases, in over 90% of the articles analyzed. Specifically, the set of articles analyzed had a cumulative prevalence of 98.1% in History and Archaeology, 93.4% in Mathematics, and 89.6% in Economics. Furthermore, social attention also increases with the number of co-authors in most cases, to above 80% in the analyzed articles. Specifically, the set of articles analyzed had a cumulative prevalence of 98.1% in History and Archaeology, 97.7% in Mathematics, and 79.7% in Economics.

Regarding the magnitude of the comparative advantage by co-authorship, the average per co-author, obtained by dividing the mean by the number of authors, is not a good measure of the increased impact due to co-authorship. This is because it penalizes articles with more than one author. That is, it tends to decrease even when the marginal contribution of an additional co-author is positive.

For this reason, we proposed a marginal analysis to measure the co-authorship effect. We defined the 'Marginal contribution' of a new co-author $n$, in the following way:

$$\text{Marginal\_contribution}(n) = \frac{\text{Mean}(n) - \text{Mean}(n-1)}{\text{Mean}(n-1)} \times 100, \quad n > 1. \tag{1}$$

A positive result in (1) represents an increase with respect to the previous value (for $n-1$), which is interpreted as a positive contribution attributable to the additional co-author. The last author incorporated into the collaboration is also known as the marginal co-author. On the contrary, a negative marginal contribution implies a reduction with respect to the previous value, which means that there is no improvement in the impact attributable to the collaboration of the last author incorporated, that is the marginal co-author.

Notice, once the first negative value in the marginal contribution is obtained in the series, the following values cease to make sense. This is because the marginal contribution is calculated with respect to the immediately previous value in the series, and not with respect to the maximum value. Therefore, the marginal contribution can provide positive increases due to previous reductions, and is not determined by real increases in the impact.

It is observed in Table 2, in the case of citations, that the marginal contribution is reduced as the number of co-authors increases. This means that, although the collaboration of an additional co-author has a positive effect on citation, its contribution becomes smaller as the number of co-authors increases. The marginal contribution in social attention also tends to decline in Economics and History and Archaeology, but there is great variability in Mathematics. Moreover, the largest marginal contributions in impact due to co-authorship are seen in History and Archaeology, both in citation and social attention.

We next analyzed the co-authorship effect on citation and social attention according to the access type (see Tables A1–A3 in Appendix A). There are important differences in average citation and social attention regarding the type of access. However, these differences are attributable to the characteristics of the access type. As can be seen, the co-authorship effect remains unchanged with respect to the aggregated data in Table 2. That is, the results obtained on co-authorship for the data disaggregated by access types are similar to those obtained for the aggregated data. According to this, the citation and social attention averages tend to increase with the number of co-authors for those co-authorships with higher prevalence.

In relation to the prevalence of the number of co-authors, it can be seen in Tables A1–A3 in the Appendix A, that the number of co-authors does not influence the decision on the type of access provided to the publication. That is, in general, the order of prevalence does not vary between types of access. Although the order of prevalence among the four dominant ones varies with the field, this order is maintained between types of access. Thus, both in Economics and in History and Archaeology, the order of prevalence in the number of authors is consecutively from 1 to 4 for most types of access (four out of five). The only exception corresponds to the OA Green. However, in Mathematics, the order of prevalence in the number of authors is 2, 3, 1, and 4 for most types of access (four out of five). The only exception in this case corresponds to the OA Bronze.

Many research articles in this study have not been cited or have not received any attention in social media (compiled in the database after four years). The proportion of articles with citations and social attention is just 24% in Economics, 12.1% in History and Archaeology, and 19.6% in Mathematics (Table 3). These low proportions correspond to many zeros in the citation and social attention distributions, which are highly skewed towards zero. For this reason, we analyzed the subgroup of articles with an impact (strictly greater than zero) both in citations and in social attention (Table 3). However, no notable differences were observed with respect to Table 1 in the linear correlation coefficients. That is, there is also no correlation between academic citation and social attention in the subgroup of articles with impact scores of greater than zero.

**Table 3.** Prevalence and Pearson correlation coefficient between times cited and social attention score for articles with impact (times cited >0 and social attention score >0).

| Number of Authors | Economics | | | History & Archaeology | | | Mathematics | | |
|---|---|---|---|---|---|---|---|---|---|
| | Np | % (of N) | Coeff. | Np | % (of N) | Coeff. | Np | % (of N) | Coeff. |
| 1 | 3014 | 17.7% | 0.38 | 4167 | 8.6% | 0.30 | 3794 | 15.1% | 0.13 |
| 2 | 4189 | 25.8% | 0.19 | 1187 | 19.8% | 0.31 | 6860 | 18.2% | 0.06 |
| 3 | 3433 | 27.5% | 0.29 | 633 | 26.4% | 0.27 | 6240 | 20.2% | 0.28 |
| 4 | 1581 | 27.9% | 0.28 | 440 | 32.9% | 0.16 | 3622 | 20.9% | 0.14 |
| 5 | 684 | 26.2% | 0.14 | 314 | 40.7% | 0.39 | 1751 | 22.4% | 0.17 |
| 6 | 294 | 21.4% | 0.21 | 204 | 45.5% | 0.75 | 917 | 24.3% | 0.05 |
| 7 | 201 | 26.2% | 0.24 | 145 | 52.5% | 0.72 | 494 | 28.4% | 0.43 |
| 8 | 138 | 27.8% | 0.76 | 102 | 54.8% | 0.46 | 311 | 30.8% | 0.45 |
| 9 | 48 * | 24.1% | 0.09 | 62 * | 56.9% | 0.76 | 210 * | 35.3% | 0.06 |
| ≥10 | 183 | 34.7% | 0.62 | 51 * | 51.5% | 0.58 | 680 | 51.7% | 0.36 |
| Total | 13,765 | 24.0% | 0.26 | 7305 | 12.1% | 0.31 | 24,879 | 19.6% | 0.14 |

\* Prevalence < 1% (Np < 1% of Total).

Regarding the prevalence of articles with impact, and its possible relationship with the number of co-authors, it can be observed how the prevalence increases with the number of co-authors in two of the three fields considered, namely History and Archaeology and Mathematics.

As can be seen in Figure 1, even within the subgroup of articles with citation and social attention, the distributions remain highly skewed, especially in terms of social attention. Notice that the mean, represented by a cross, is much higher than the median, represented by the central line in the box. This means that a small proportion of articles accumulates a large part of the citations and, above all, most of the social attention.

Figure 1 also shows how both citation and social attention in general grow with the number of co-authors. The differences between groups are all significant ($p < 0.01$). This trend is observed both in the mean and in the median, and even in the rest of the quartiles in the distribution represented by the boxes and whiskers in the figure.

Moreover, in Figure 1, the differences in the orders of magnitude in the impacts between fields can be observed. Thus, articles in History and Archaeology tend to receive fewer academic citations than those in the other two fields (around half on average). However, the social attention for articles in History and Archaeology is similar to that for articles in Economics, and far superior to those in Mathematics.

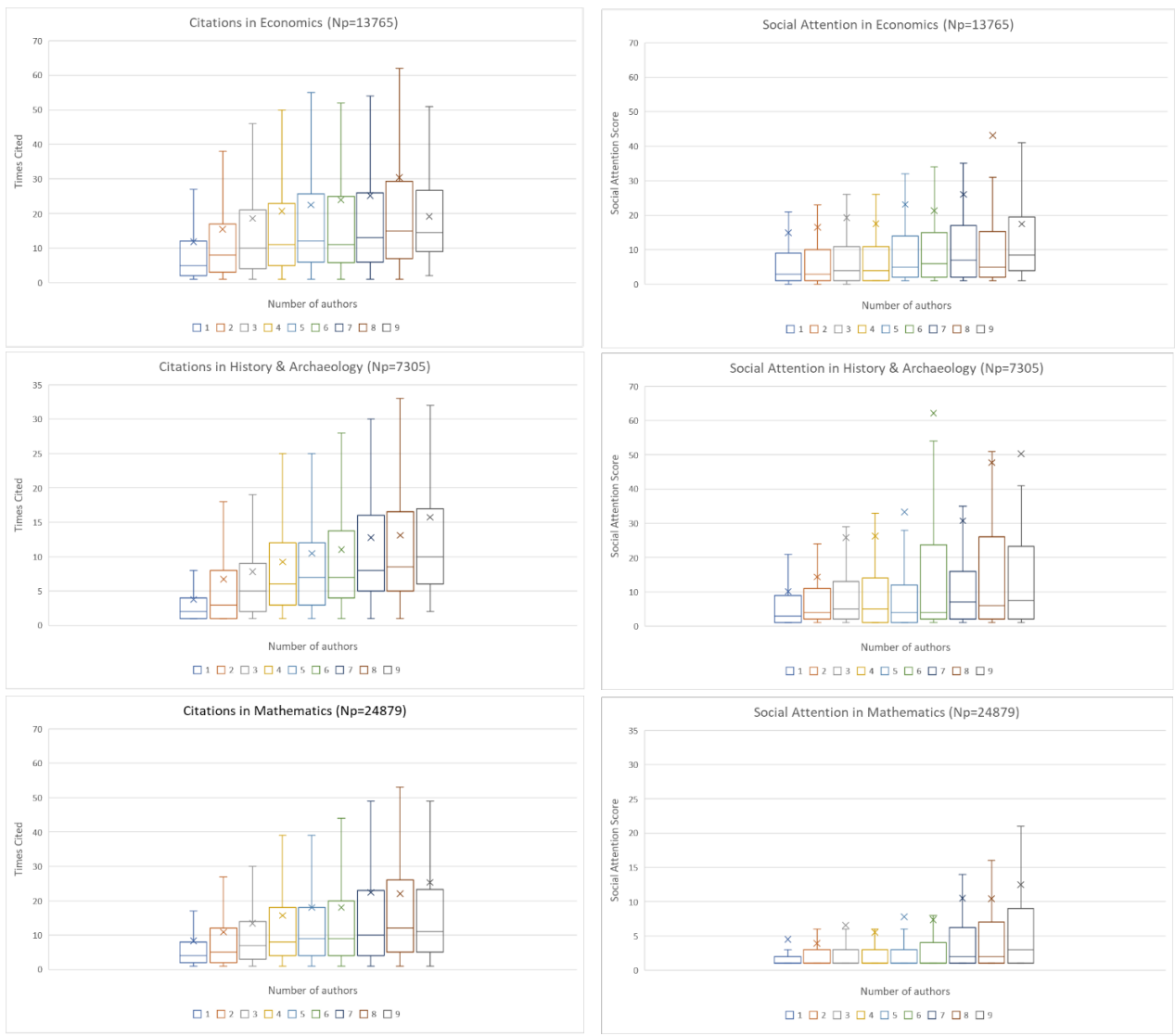

**Figure 1.** Times cited and social attention score from one to nine authors. The mean is represented by a cross. The differences between groups are all significant ($p < 0.01$).

Finally, we analyzed the title length of the articles, in terms of the number of characters, in the total sample (Figure 2). The differences between groups of co-authorships are all significant in Mathematics ($p < 0.01$). In the other two disciplines, there are significant differences between co-authorships ($p < 0.01$) except for the groups 6 to 9 in Economics, and the groups 4 to 9 in History and Archaeology. It should be noted that the differences between groups are significant in the case of the most prevalent co-authorships in Economics and History and Archaeology. It is also noteworthy that collaborations with more than six authors in Economics or with more than four authors in History and Archaeology are rare (Table 2).

As can be seen in Figure 2, there is a positive relationship, for up to six authors, between the number of authors and the title length. The greater the number of co-authors, the greater the length of the title.

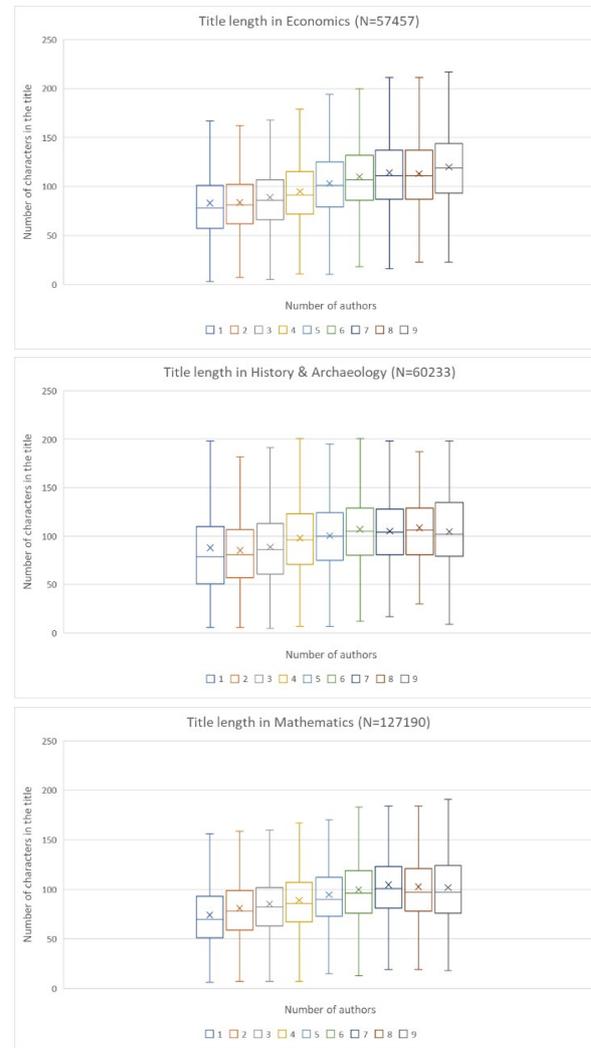

**Figure 2.** Title length from one to nine authors. The mean is represented by a cross. The differences between groups are all significant in Mathematics ($p < 0.01$). In the case of Economics, there are significant differences between co-authorships ($p < 0.01$) except in the groups 6 to 9. Finally, in History & Archaeology, the differences are significant between groups ($p < 0.01$) except for co-authorships 4 to 9.

However, the title length does not correlate with the citations received, with coefficients in terms of absolute value of lower than 0.20 (Table 4). There are negative signs in Economics (less than 0.20 in absolute value), positive signs in History and Archaeology (in absolute value below 0.14), and signs vary in Mathematics (below 0.09 in absolute value).

A similar trend was observed for social attention. The extension in the title does not correlate with social attention, with coefficients below 0.24 in absolute value. Nevertheless, it should be noted that this is typical for all cases with negative signs.

**Table 4.** Pearson correlation coefficient between title length and variables, including times cited and social attention score in relation to co-authorship.

| Number of Authors | Economics | | History & Archaeology | | Mathematics | |
|---|---|---|---|---|---|---|
| | Times Cited | Social Attention Score | Times Cited | Social Attention Score | Times Cited | Social Attention Score |
| 1 | −0.04 | −0.03 | −0.04 | −0.03 | 0.04 | −0.04 |
| 2 | −0.03 | −0.03 | 0.03 | −0.01 | 0.05 | −0.02 |
| 3 | −0.05 | −0.03 | 0.10 | 0.00 | 0.02 | −0.02 |
| 4 | −0.07 | −0.04 | 0.10 | −0.02 | 0.02 | −0.02 |
| 5 | −0.09 | −0.06 | 0.03 | −0.03 | −0.02 | −0.02 |
| 6 | −0.10 | −0.07 | 0.06 | −0.07 | 0.01 | −0.03 |
| 7 | −0.11 | −0.05 | 0.14 | −0.01 | −0.02 | −0.01 |
| 8 | −0.08 | −0.10 | 0.00 | −0.10 | −0.09 | −0.10 |
| 9 | −0.20 | −0.24 | 0.03 | 0.00 | −0.03 | −0.01 |
| ≥10 | −0.29 | −0.22 | 0.00 | −0.11 | −0.07 | −0.05 |

## 4. Discussion

Academic citation and social attention are influenced by many factors. Among these factors are the research field and co-authorship. As evidenced, citation and social attention do not correlate with each other and, therefore, they measure different dimensions in the impact of research results. This is something that has already been pointed out in the literature for unweighted indicators [23,24].

In terms of the number of co-authors in Economics, the most frequent is a single author (30%), while 80% of the articles have less than three authors. This confirms the results obtained by [29] using data from more than 700,000 journal articles. However, the trend in the last three decades is towards a reduction in the number of articles with a single author [30].

As evidenced, research in Mathematics is the field that generates the least social attention among the three disciplines analyzed. Note that Mathematics is not characterized as a media discipline [31]. Although with notable exceptions, mathematical research is frequently basic, and its social and economic applications usually become evident in the long term. In addition, mathematical discoveries often provide methods and tools that are later applied in other disciplines. At the same time, mathematical results are frequently communicated exclusively through specialized journals, with less use of platforms and social networks. This usually generates less discussion of mathematical results in society compared to other more media-focused areas, such as History and Archaeology, for example.

Our results indicate that the marginal contribution reduces as the number of co-authors increases. This means that, although the effect of the collaboration of an additional co-author is in general positive on citation and social attention, its contribution becomes less pronounced as the number of co-authors increases. Some explanations for these results can be given from a theoretical point of view. The marginal effect of an additional author probabilistically implies more citations and social attention. This may be due to two main factors. On the one hand, the contribution of new knowledge that generates innovative ideas may have a positive impact [1]. However, it can also be influenced by the possible self-mentions of the marginal author, that is, the auto-cites and social debate may be generated by the marginal author through their social networks.

As indicated, an additional author can self-cite the paper in their subsequent publications and communicate the results obtained through their social networks. We can argue why in this case the contribution to the generation of impact is always positive. When authorship increases from a single author to two, for example, both the self-citations and the social self-discussion of the authors would be multiplied by two (a marginal increase of 100%). When increasing the number from two to three authors, self-citations and social self-discussion are multiplied by 1.5 (a marginal increase of 50%). When increasing the authors from three to four, both own impacts would be multiplied by 1.33 (a marginal increase of 33%), and so on. As can be seen, although the effect of the own impact is progressively reduced, the marginal multiplication factor is greater than one, with a positive marginal increase.

In general, when increasing from $n$ to $n + 1$ authors, the part of the impact associated with the action of the authors themselves, both in self-citations and in self-generation of social attention, is multiplied by a factor of $1 + (1/n)$, where $n \geq 1$. This means that the marginal effect of self-mention is theoretically $(1/n) \times 100\%$, where $n \geq 1$ denotes the number of authors.

For this effect of theoretical self-mention, the effect of generating innovative ideas should also be mentioned, which could only be estimated empirically.

In relation to the co-authorship effect on citation and social attention according to the access type, we observed that, in general, the order of prevalence does not vary between types of access. The only exception in Economics and History and Archaeology corresponds to the OA Green. This may be because in this modality it is enough for any of the co-authors to submit the document in an open access repository. However, in Mathematics, the only exception corresponds to the OA Bronze. Note that the decision on this modality corresponds to the publisher and not the authors.

Many research articles in this study have not been cited or have not received any attention in social media (compiled in the database after four years). This is something that is observed frequently in the literature [32].

As evidenced, the prevalence of articles with impact increases with the number of co-authors in two of the three fields considered, in History and Archaeology and Mathematics. This means that the greater the number of co-authors, the greater the probability of being cited in academic articles and mentioned on social media platforms.

We analyzed the title length of the articles. Note that the length of the title can be considered a proxy to the specificity/generality of the article topic. In general, long titles correspond to more specific topics while short titles correspond to more general topics that may be easier to understand, and hence attract more citations [33]. We obtained a positive relationship, for up to six authors, between the number of authors and the title length. This may be due to the fact that research collaboration tends to be more specific in relation to the topic and, therefore, they tend to have longer titles.

However, the title length does not correlate with the citations received. The negative signs in the results in Economics are related to previous studies [29,34,35]. Some authors observed negative correlations between the title length of an economics article and citations [34,35]. Furthermore, other authors provide evidence on how the authorship of papers is related to the number of citations and the length of the title [29].

In relation to the results obtained, some considerations should be made. The drivers for a positive marginal effect on the impact, as the number of collaborators increases, can be multiple. We pointed out two main factors, the generation of innovative ideas based on the knowledge of collaborators [1], and the self-citations and self-mentions on social networks. Beyond the distribution of tasks, the collaborative articles try to take advantage of the complementarity in the specializations of the collaborators. This often leads to innovative ideas that facilitate their publication in high-impact journals. These journals tend to have higher visibility and a wider audience, so they tend to attract more citations.

In this context, we incorporated the impact of the journals into the dataset. We used the *CiteScore* provided by the *Scopus* database for 2018, the year of publication for the analyzed articles. Note that the *Dimensions.ai* database does not provide impact indicators at the journal level. However, by volume of journals with calculated impact, both databases have a similar order of magnitude, which is much higher than the *JCR* in *Web of Science*.

We obtained a low correlation between the journal impact and the citations of the articles in Economics (less than 0.3 for most co-authorships), and no correlation in the other two disciplines analyzed (below 0.1 for all co-authorships). In all cases, they were lower than those previously obtained between citation and social attention. Thus, the average impact of the journal is a very poor predictor of the real impact achieved by the articles it publishes. This is well recognized in the literature, and it is because high-impact journals typically receive a large proportion of their citations from a small number of high-impact papers. This is why it is recommended, in research assessments, to use the real impact of the papers instead of the journal impact factor when evaluating the performance of researchers.

Given this low correlation between the average impact of the journal and the individual impact of each paper, it is not surprising that the correlations between *CiteScore* and social attention are even lower, below 0.1 in terms of absolute value for most of the co-authorships.

It should be noted that in addition to regular cooperation that increases the number of co-authors, there are cases of misuse, such as adding names that did not actually participate, but their reputation in the field can draw attention [36].

Some considerations can be made about using hybrid indicators. This study used the 'altmetric attention score'. This is a hybrid indicator that aggregates multiple sources into a single score [28]. Hybrid indicators cannot be robustly interpreted. For this reason, it should be avoided when evaluating researchers, especially in recruitment processes and internal promotions. However, in this work, this indicator was used to evaluate the research process itself and not the researchers.

Altmetric indicators have the advantage of measuring different types of impacts beyond academic citations. They also have the potential to capture earlier impact evidence. This is useful in self-evaluations. This is also useful when studying science itself, as is the case in this study. Nevertheless, social attention must be used cautiously because it could provide a partial and biased view of all types of social impact.

## 5. Conclusions

We quantified the contribution of collaboration to the academic citation and social attention of research. As a novel methodology, we measured this contribution due to co-authorship through a marginal analysis, i.e., through the contribution to the impact attributable to an additional co-author. We analyzed for the first time the possible effect of the access type to the publication, particularly open access modalities.

The results in this research, which assumes co-authorship as a proxy of research collaboration, show that citation and social attention do not correlate. Therefore, they measure different dimensions in the impact of research results.

We observed comparative advantages due to collaboration both in citation and social attention. Thus, we obtained comparative advantages in Economics for up to four authors in citations and up to three authors in social attention. In History and Archaeology, comparative advantages were observed for up to five authors both in citations and social attention, after which collaborations cease to be frequent. In Mathematics, comparative advantages were obtained for up to five authors in citations and up to seven authors in social attention.

Citations increased with the number of co-authors in 98% of articles in History and Archaeology, 93% of articles in Mathematics, and 90% of articles in Economics. Furthermore, social attention also increased with the number of co-authors in 98% of articles in History and Archaeology and Mathematics, and 80% of articles in Economics. However, the marginal contribution of collaboration tends to decline in citation and social attention.

Although we identified important differences in average citation and social attention regarding the type of access to the publication, these differences are attributable exclusively to the characteristics of the access type.

A small proportion of articles accumulates a large part of the citations and, above all, most of the social attention. Four years following their publication, the proportion of articles with citations and social attention is just 24% in Economics, 12% in History and Archaeology, and 20% in Mathematics.

There are differences in the orders of magnitude in the impacts between fields. Thus, articles in History and Archaeology tend to receive fewer academic citations than those in the other two fields (around half on average). However, the social attention to articles in History and Archaeology is similar to articles in Economics, and far superior to those in Mathematics.

Finally, the length of the article title, which can be considered as a proxy for the specificity of its content, increases with the number of co-authors. However, it does not correlate with citations or with social attention.

**Author Contributions:** Conceptualization, P.D.-G.; methodology, P.D.-G. and M.I.D.-G.; software, M.I.D.-G.; validation, P.D.-G. and M.I.D.-G.; formal analysis, P.D.-G. and M.I.D.-G.; investigation, P.D.-G. and M.I.D.-G.; resources, P.D.-G.; data curation, P.D.-G. and M.I.D.-G.; writing—original draft preparation, P.D.-G. and M.I.D.-G.; writing—review and editing, P.D.-G. and M.I.D.-G.; supervision, P.D.-G. All authors have read and agreed to the published version of the manuscript.

**Funding:** This research received no external funding.

**Institutional Review Board Statement:** Not applicable.

**Informed Consent Statement:** Not applicable.

**Data Availability Statement:** Data sourced from Dimensions.ai, an inter-linked research information system provided by Digital Science (accessed on 15 March 2022).

**Acknowledgments:** Access to advanced features of the database allowed by *Digital Science*.

**Conflicts of Interest:** The authors declare no conflict of interest.

## Appendix A

See Tables A1–A3.

**Table A1.** Times cited and social attention score in Economics in relation to the access type and co-authorship.

| Access Type | Number of Authors | Articles | | Times Cited | | | Social Attention Score | | |
|---|---|---|---|---|---|---|---|---|---|
| | | Number N | Prevalence | Mean | Mean per Author | Marginal Contribution | Mean | Mean per Author | Marginal Contribution |
| OA Gold | 1 | 4796 | 32.4% | 1.7 | 1.7 | - | 1.1 | 1.1 | - |
| | 2 | 4227 | 28.5% | 2.8 | 1.4 | 71% | 2.1 | 1.0 | 98% |
| | 3 | 3335 | 22.5% | 3.4 | 1.1 | 20% | 3.4 | 1.1 | 63% |
| | 4 | 1342 | 9.1% | 4.9 | 1.2 | 44% | 2.8 | 0.7 | −18% |
| | 5 | 532 | 3.6% | 6.9 | 1.4 | 42% | 7.3 | 1.5 | 163% |
| | 6 | 254 | 1.7% | 7.0 | 1.2 | 1% | 4.7 | 0.8 | −35% |
| | 7 * | 132 | 0.9% | 12.1 | 1.7 | 72% | 23.3 | 3.3 | 394% |
| | 8 * | 68 | 0.5% | 11.4 | 1.4 | −6% | 26.2 | 3.3 | 13% |
| | 9 * | 34 | 0.2% | 8.2 | 0.9 | −28% | 9.3 | 1.0 | −65% |
| | ≥10 * | 102 | 0.7% | 17.8 | - | 117% | 28.2 | - | 205% |
| | Total | 14,822 | | | | | | | |
| OA Hybrid | 1 | 932 | 29.1% | 4.0 | 4.0 | - | 6.2 | 6.2 | - |
| | 2 | 899 | 28.1% | 8.2 | 4.1 | 104% | 8.9 | 4.4 | 42% |
| | 3 | 740 | 23.1% | 13.4 | 4.5 | 63% | 10.1 | 3.4 | 14% |
| | 4 | 305 | 9.5% | 15.9 | 4.0 | 19% | 10.1 | 2.5 | 0% |
| | 5 | 134 | 4.2% | 20.6 | 4.1 | 30% | 21.3 | 4.3 | 111% |
| | 6 | 69 | 2.2% | 25.0 | 4.2 | 21% | 20.5 | 3.4 | −4% |
|  | 7 | 42 | 1.3% | 28.8 | 4.1 | 15% | 21.3 | 3.0 | 4% |
|  | 8 | 33 | 1.0% | 61.1 | 7.6 | 112% | 55.9 | 7.0 | 162% |
|  | 9 * | 8 | 0.2% | 28.5 | 3.2 | −53% | 16.1 | 1.8 | −71% |
|  | ≥10 | 39 | 1.2% | 43.1 | - | 51% | 34.4 | - | 113% |
|  | Total | 3201 |  |  |  |  |  |  |  |
| OA Green | 1 | 1634 | 19.9% | 9.0 | 9.0 | - | 6.3 | 6.3 | - |
|  | 2 | 2835 | 34.6% | 11.1 | 5.6 | 24% | 7.0 | 3.5 | 11% |
|  | 3 | 2303 | 28.1% | 13.9 | 4.6 | 25% | 8.2 | 2.7 | 17% |
|  | 4 | 862 | 10.5% | 20.1 | 5.0 | 44% | 7.6 | 1.9 | −6% |
|  | 5 | 286 | 3.5% | 19.1 | 3.8 | −5% | 16.6 | 3.3 | 118% |
|  | 6 | 104 | 1.3% | 23.9 | 4.0 | 25% | 8.2 | 1.4 | −51% |
|  | 7 * | 67 | 0.8% | 25.4 | 3.6 | 7% | 9.1 | 1.3 | 10% |
|  | 8 * | 40 | 0.5% | 17.9 | 2.2 | −30% | 17.6 | 2.2 | 94% |
|  | 9 * | 13 | 0.2% | 26.8 | 3.0 | 50% | 10.5 | 1.2 | −40% |
|  | ≥10 * | 49 | 0.6% | 54.3 | - | 102% | 67.2 | - | 538% |
|  | Total | 8193 |  |  |  |  |  |  |  |
| OA Bronze | 1 | 1641 | 20.0% | 2.6 | 2.6 | - | 4.9 | 4.9 | - |
|  | 2 | 1526 | 18.6% | 4.5 | 2.3 | 75% | 8.2 | 4.1 | 66% |
|  | 3 | 1482 | 18.1% | 5.0 | 1.7 | 10% | 7.3 | 2.4 | −11% |
|  | 4 | 1119 | 13.6% | 3.3 | 0.8 | −34% | 3.4 | 0.9 | −53% |
|  | 5 | 856 | 10.4% | 1.3 | 0.3 | −59% | 1.2 | 0.2 | −65% |
|  | 6 | 599 | 7.3% | 1.6 | 0.3 | 20% | 1.4 | 0.2 | 18% |
|  | 7 | 361 | 4.4% | 1.5 | 0.2 | −8% | 0.8 | 0.1 | −45% |
|  | 8 | 256 | 3.1% | 1.0 | 0.1 | −31% | 3.0 | 0.4 | 285% |
|  | 9 | 110 | 1.3% | 1.2 | 0.1 | 14% | 1.6 | 0.2 | −48% |
|  | ≥10 | 260 | 3.2% | 3.1 | - | 169% | 1.8 | - | 14% |
|  | Total | 8210 |  |  |  |  |  |  |  |
| Closed | 1 | 8069 | 35.0% | 3.6 | 3.6 | - | 2.5 | 2.5 | - |
|  | 2 | 6776 | 29.4% | 7.9 | 3.9 | 120% | 3.6 | 1.8 | 47% |
|  | 3 | 4606 | 20.0% | 10.3 | 3.4 | 31% | 4.1 | 1.4 | 14% |
|  | 4 | 2048 | 8.9% | 13.0 | 3.2 | 26% | 5.3 | 1.3 | 29% |
|  | 5 | 807 | 3.5% | 16.1 | 3.2 | 24% | 4.3 | 0.9 | −18% |
|  | 6 | 351 | 1.5% | 17.6 | 2.9 | 9% | 6.6 | 1.1 | 53% |
|  | 7 * | 165 | 0.7% | 21.9 | 3.1 | 24% | 2.8 | 0.4 | −58% |
|  | 8 * | 99 | 0.4% | 20.3 | 2.5 | −7% | 9.6 | 1.2 | 241% |
|  | 9 * | 34 | 0.1% | 13.4 | 1.5 | −34% | 3.6 | 0.4 | −63% |
|  | ≥10 * | 76 | 0.3% | 23.6 | - | 76% | 19.6 | - | 452% |
|  | Total | 23,031 |  |  |  |  |  |  |  |

* Prevalence < 1% (N < 1% of Total).

**Table A2.** Times cited and social attention score in History & Archaeology in relation to the access type and co-authorship.

| Access Type | Number of Authors | Articles | | Times Cited | | | Social Attention Score | | |
|---|---|---|---|---|---|---|---|---|---|
| | | Number N | Prevalence | Mean | Mean per Author | Marginal Contribution | Mean | Mean per Author | Marginal Contribution |
| OA Gold | 1 | 10,710 | 76.2% | 0.4 | 0.4 | - | 0.6 | 0.6 | - |
| | 2 | 1956 | 13.9% | 0.8 | 0.4 | 108% | 1.0 | 0.5 | 56% |
| | 3 * | 714 | 5.1% | 2.0 | 0.7 | 141% | 8.5 | 2.8 | 775% |
| | 4 * | 323 | 2.3% | 1.7 | 0.4 | −11% | 4.0 | 1.0 | −53% |
| | 5 * | 167 | 1.2% | 4.1 | 0.8 | 139% | 12.0 | 2.4 | 199% |
| | 6 * | 79 | 0.6% | 5.9 | 1.0 | 43% | 33.8 | 5.6 | 183% |
| | 7 * | 41 | 0.3% | 9.5 | 1.4 | 61% | 32.4 | 4.6 | −4% |
| | 8 * | 27 | 0.2% | 12.0 | 1.5 | 26% | 94.4 | 11.8 | 191% |
| | 9 * | 15 | 0.1% | 9.2 | 1.0 | −23% | 25.7 | 2.9 | −73% |
| | ≥10 * | 17 | 0.1% | 20.9 | - | 127% | 211.8 | - | 725% |
| | Total | 14,049 | | | | | | | |
| OA Hybrid | 1 | 1703 | 67.7% | 1.0 | 1.0 | - | 2.0 | 2.0 | - |
| | 2 | 411 | 16.3% | 2.5 | 1.3 | 159% | 5.8 | 2.9 | 191% |
| | 3 | 132 | 5.3% | 3.9 | 1.3 | 56% | 15.9 | 5.3 | 174% |
| | 4 | 127 | 5.1% | 6.6 | 1.7 | 69% | 26.5 | 6.6 | 66% |
| | 5 | 51 | 2.0% | 11.0 | 2.2 | 66% | 82.1 | 16.4 | 209% |
| | 6 | 29 | 1.2% | 7.4 | 1.2 | −33% | 12.9 | 2.2 | −84% |
| | 7 * | 24 | 1.0% | 15.0 | 2.1 | 104% | 37.3 | 5.3 | 189% |
| | 8 * | 21 | 0.8% | 8.3 | 1.0 | −44% | 10.8 | 1.4 | −71% |
| | 9 * | 4 | 0.2% | 9.5 | 1.1 | 14% | 54.5 | 6.1 | 405% |
| | ≥10 * | 12 | 0.5% | 24.1 | - | 153% | 100.1 | - | 84% |
| | Total | 2514 | | | | | | | |
| OA Green | 1 | 2348 | 64.0% | 1.4 | 1.4 | - | 3.9 | 3.9 | - |
| | 2 | 450 | 12.3% | 4.2 | 2.1 | 194% | 5.4 | 2.7 | 39% |
| | 3 | 242 | 6.6% | 5.8 | 1.9 | 38% | 12.9 | 4.3 | 138% |
| | 4 | 323 | 8.8% | 8.4 | 2.1 | 46% | 10.5 | 2.6 | −19% |
| | 5 | 97 | 2.6% | 8.4 | 1.7 | 0% | 12.7 | 2.5 | 21% |
| | 6 | 70 | 1.9% | 10.2 | 1.7 | 22% | 56.9 | 9.5 | 347% |
| | 7 | 56 | 1.5% | 10.4 | 1.5 | 1% | 16.2 | 2.3 | −72% |
| | 8 * | 32 | 0.9% | 7.9 | 1.0 | −24% | 26.8 | 3.3 | 66% |
| | 9 * | 22 | 0.6% | 11.1 | 1.2 | 40% | 11.1 | 1.2 | −59% |
| | ≥10 * | 29 | 0.8% | 9.1 | - | −18% | 9.1 | - | −18% |
| | Total | 3669 | | | | | | | |
| OA Bronze | 1 | 3411 | 74.6% | 0.7 | 0.7 | - | 2.0 | 2.0 | - |
| | 2 | 688 | 15.0% | 2.0 | 1.0 | 179% | 6.0 | 3.0 | 209% |
| | 3 | 246 | 5.4% | 3.2 | 1.1 | 55% | 7.9 | 2.6 | 30% |
| | 4 | 115 | 2.5% | 3.8 | 0.9 | 20% | 7.3 | 1.8 | −7% |
| | 5 | 51 | 1.1% | 4.8 | 1.0 | 26% | 15.2 | 3.0 | 108% |
| | 6 * | 26 | 0.6% | 7.8 | 1.3 | 63% | 64.0 | 10.7 | 320% |
| | 7 * | 14 | 0.3% | 7.1 | 1.0 | −9% | 23.6 | 3.4 | −63% |
| | 8 * | 14 | 0.3% | 13.4 | 1.7 | 89% | 58.8 | 7.3 | 149% |
| | 9 * | 6 | 0.1% | 40.0 | 4.4 | 200% | 171.3 | 19.0 | 191% |
| | ≥10 * | 3 | 0.1% | 12.0 | - | −70% | 61.7 | - | −64% |
|  | Total | 4574 |  |  |  |  |  |  |  |
| Closed | 1 | 30,282 | 85.5% | 0.5 | 0.5 | - | 1.2 | 1.2 | - |
|  | 2 | 2483 | 7.0% | 2.5 | 1.2 | 370% | 3.5 | 1.8 | 200% |
|  | 3 | 1065 | 3.0% | 3.6 | 1.2 | 43% | 4.0 | 1.3 | 13% |
|  | 4 | 614 | 1.7% | 5.0 | 1.2 | 39% | 7.7 | 1.9 | 93% |
|  | 5 | 406 | 1.1% | 5.6 | 1.1 | 12% | 6.3 | 1.3 | −19% |
|  | 6 * | 244 | 0.7% | 6.6 | 1.1 | 18% | 17.0 | 2.8 | 172% |
|  | 7 * | 141 | 0.4% | 7.7 | 1.1 | 17% | 7.6 | 1.1 | −56% |
|  | 8 * | 92 | 0.3% | 8.9 | 1.1 | 16% | 4.5 | 0.6 | −40% |
|  | 9 * | 62 | 0.2% | 8.6 | 1.0 | −3% | 20.2 | 2.2 | 345% |
|  | ≥10 * | 38 | 0.1% | 6.5 | - | −24% | 8.4 | - | −59% |
|  | Total | 35,427 |  |  |  |  |  |  |  |

* Prevalence < 1% (N < 1% of Total).

**Table A3.** Times cited and social attention score in Mathematics in relation to the access type and co-authorship.

| Access Type | Number of Authors | Articles | | Times Cited | | | Social Attention Score | | |
|---|---|---|---|---|---|---|---|---|---|
|  |  | Number N | Prevalence | Mean | Mean per Author | Marginal Contribution | Mean | Mean per Author | Marginal Contribution |
| OA Gold | 1 | 4532 | 16.3% | 3.8 | 3.8 | - | 0.7 | 0.7 | - |
|  | 2 | 7200 | 25.9% | 5.2 | 2.6 | 39% | 1.4 | 0.7 | 92% |
|  | 3 | 6947 | 25.0% | 6.8 | 2.3 | 29% | 1.2 | 0.4 | −14% |
|  | 4 | 4405 | 15.8% | 7.9 | 2.0 | 17% | 0.4 | 0.1 | −64% |
|  | 5 | 2342 | 8.4% | 8.7 | 1.7 | 10% | 1.5 | 0.3 | 258% |
|  | 6 | 1179 | 4.2% | 9.7 | 1.6 | 12% | 1.3 | 0.2 | −18% |
|  | 7 | 528 | 1.9% | 10.9 | 1.6 | 12% | 3.0 | 0.4 | 137% |
|  | 8 | 290 | 1.0% | 10.1 | 1.3 | −7% | 2.5 | 0.3 | −16% |
|  | 9 * | 108 | 0.4% | 8.7 | 1.0 | −15% | 1.3 | 0.1 | −49% |
|  | ≥10 | 303 | 1.1% | 15.0 | - | 73% | 7.5 | - | 491% |
|  | Total | 27,834 |  |  |  |  |  |  |  |
| OA Hybrid | 1 | 1059 | 18.2% | 4.3 | 4.3 | - | 1.5 | 1.5 | - |
|  | 2 | 1780 | 30.6% | 7.3 | 3.7 | 69% | 1.7 | 0.8 | 13% |
|  | 3 | 1391 | 23.9% | 11.7 | 3.9 | 60% | 4.2 | 1.4 | 149% |
|  | 4 | 753 | 13.0% | 12.2 | 3.1 | 5% | 3.5 | 0.9 | −16% |
|  | 5 | 365 | 6.3% | 16.8 | 3.4 | 37% | 7.6 | 1.5 | 116% |
|  | 6 | 154 | 2.6% | 15.8 | 2.6 | −6% | 6.4 | 1.1 | −16% |
|  | 7 | 73 | 1.3% | 22.3 | 3.2 | 41% | 15.6 | 2.2 | 144% |
|  | 8 | 66 | 1.1% | 28.9 | 3.6 | 30% | 8.9 | 1.1 | −43% |
|  | 9 * | 50 | 0.9% | 19.0 | 2.1 | −34% | 13.5 | 1.5 | 52% |
|  | ≥10 | 123 | 2.1% | 32.6 | - | 72% | 27.4 | - | 103% |
|  | Total | 5814 |  |  |  |  |  |  |  |
| OA Green | 1 | 5563 | 20.8% | 4.0 | 4.0 | - | 1.1 | 1.1 | - |
|  | 2 | 9029 | 33.8% | 6.7 | 3.3 | 67% | 1.1 | 0.5 | 0% |
|  | 3 | 6751 | 25.3% | 8.7 | 2.9 | 31% | 1.8 | 0.6 | 74% |
|  | 4 | 3074 | 11.5% | 11.3 | 2.8 | 30% | 2.3 | 0.6 | 24% |
|  | 5 | 1159 | 4.3% | 14.7 | 2.9 | 30% | 3.2 | 0.6 | 40% |
|  | 6 | 485 | 1.8% | 14.8 | 2.5 | 1% | 2.8 | 0.5 | −13% |
|  | 7 * | 222 | 0.8% | 21.0 | 3.0 | 42% | 6.3 | 0.9 | 125% |
|  | 8 * | 120 | 0.4% | 19.1 | 2.4 | −9% | 9.3 | 1.2 | 48% |
|  | 9 * | 108 | 0.4% | 17.9 | 2.0 | −6% | 7.2 | 0.8 | −23% |
|  | ≥10 * | 185 | 0.7% | 36.9 | - | 106% | 22.9 | - | 219% |
|  | Total | 26,696 |  |  |  |  |  |  |  |
| OA Bronze | 1 | 2403 | 26.7% | 2.5 | 2.5 | - | 2.4 | 2.4 | - |
|  | 2 | 2736 | 30.4% | 4.6 | 2.3 | 81% | 2.0 | 1.0 | −16% |
|  | 3 | 1974 | 21.9% | 5.8 | 1.9 | 26% | 1.5 | 0.5 | −23% |
|  | 4 | 1023 | 11.4% | 8.0 | 2.0 | 39% | 1.1 | 0.3 | −29% |
|  | 5 | 336 | 3.7% | 7.8 | 1.6 | −3% | 1.2 | 0.2 | 15% |
|  | 6 | 198 | 2.2% | 8.3 | 1.4 | 6% | 2.5 | 0.4 | 103% |
|  | 7 * | 84 | 0.9% | 11.8 | 1.7 | 43% | 3.8 | 0.5 | 52% |
|  | 8 * | 62 | 0.7% | 9.0 | 1.1 | −24% | 1.9 | 0.2 | −50% |
|  | 9 * | 34 | 0.4% | 9.2 | 1.0 | 2% | 7.4 | 0.8 | 288% |
|  | ≥10 | 147 | 1.6% | 14.6 | - | 59% | 6.5 | - | −13% |
|  | Total | 8997 |  |  |  |  |  |  |  |
| Closed | 1 | 11,590 | 20.0% | 3.2 | 3.2 | - | 0.4 | 0.4 | - |
|  | 2 | 16,863 | 29.2% | 6.1 | 3.0 | 90% | 0.3 | 0.1 | −22% |
|  | 3 | 13,825 | 23.9% | 8.0 | 2.7 | 32% | 0.4 | 0.1 | 24% |
|  | 4 | 8052 | 13.9% | 10.4 | 2.6 | 30% | 0.4 | 0.1 | 19% |
|  | 5 | 3613 | 6.2% | 11.9 | 2.4 | 14% | 0.9 | 0.2 | 109% |
|  | 6 | 1750 | 3.0% | 10.0 | 1.7 | −16% | 1.4 | 0.2 | 59% |
|  | 7 | 832 | 1.4% | 8.8 | 1.3 | −12% | 1.0 | 0.1 | −29% |
|  | 8 * | 472 | 0.8% | 7.8 | 1.0 | −12% | 1.7 | 0.2 | 66% |
|  | 9 * | 295 | 0.5% | 12.4 | 1.4 | 60% | 2.8 | 0.3 | 64% |
|  | ≥10 | 557 | 1.0% | 9.9 | - | −20% | 5.3 | - | 92% |
|  | Total | 57,849 |  |  |  |  |  |  |  |

* Prevalence < 1% (N < 1% of Total).